\documentclass[a4paper, superscriptaddress, aps, prl, onecolumn]{revtex4}
\usepackage{amsmath}
\usepackage{graphicx}

\begin{document}
\title{Effect of material stiffness on intermodulation response in dynamic atomic force microscopy}
\author{Daniel Platz}
\email{platz@kth.se}
\author{Daniel Forchheimer}
\author{Erik A. Thol\'{e}n}
\author{Carsten Hutter}
\author{David B. Haviland}
\affiliation{Royal Institute of Technology (KTH), Section for Nanostructure Physics, Albanova University Center, SE-106 91 Stockholm, Sweden}

\begin{abstract}
We perform simulations and experiments on an oscillating atomic force microscope cantilever approaching a surface, where the intermodulation response of the cantilever driven with two pure harmonic tones is investigated.  In the simulations, the tip and surface interact with a conservative nonlinear force, and the parameter space of approach distance and surface stiffness is explored.  Approach experiments are carried out on three surfaces with widely varying stiffness.  Qualitative similarities between simulations and experiment can be seen, but quantitative comparison is difficult due to the overly idealized tip-surface force model used in the simulation.  

\end{abstract}

\maketitle

\section{Introduction}

Methods and techniques from nonlinear systems analysis have the potential to greatly enhance the surface analysis capabilities of the atomic force microscope (AFM). The nonlinearity of interest in AFM is the minute force between a very sharp tip and a surface, which depends on the material composition and geometry of the tip and surface at the nanometer scale. This nonlinear tip-surface force perturbs the linear dynamics of the freely oscillating AFM cantilever, giving rise to intermodulation, or mixing of different drive frequencies. When the AFM cantilever is driven with two pure harmonic tones at frequencies $f_1$ and $f_2$, the nonlinear tip-surface force will generate intermodulation products of the drive tones in the cantilevers response at frequencies $nf_1 + mf_2$ , where $n$ and $m$ are integers. With the appropriate choice of $f_1$ and $f_2$ , many intermodulation products of high order $\vert n \vert + \vert m \vert$ can be placed near resonance, where large transfer gain allows for detection of the response with enhanced sensitivity \cite{Platz2008}.  Thus, in contrast to traditional dynamic AFM, where response amplitude and phase are measured only at the drive frequency, intermodulation AFM acquires many (typically of order 30) amplitude and phase quantities, which together contain information about the nonlinear tip-surface force that created the intermodulation response.  By analysis of the intermodulation spectrum one can, in principal, reconstruct a polynomial approximation to a conservative tip-surface force $F_{ts}(z)$ as a function of the cantilever displacement $z$ \cite{Hutter_PRL}.  Reconstruction methods which include the possibility of non-conservative tip-surface forces are under development. 

Reconstructing the tip-surface force from analysis of the nonlinear cantilever dynamics is one of the current trends in AFM research \cite{Garcia2002197,PhysRevLett.97.036104,0957-4484-19-37-375704,sahin_nn07,MartinStark06252002}.  This approach has historical roots:  The analytical power of the AFM has thus far been defined in terms of the instruments ability to measure force-distance curves $F_{ts}(z)$ by monitoring the {\em static} deflection of the cantilever while slowly approaching the surface.  It is natural to extend the interpretation of approach curves in { \em dynamic} AFM in terms of force-distance curves. However, this is perhaps not the best method of analyzing the nonlinear dynamics of a driven cantilever impacting on a surface.  The problem can become quite complex if many eigenmodes of the cantilever and the fast feedback used in AFM to track the surface, are all accounted for in a full description of the dynamical system.  If the goal of dynamic methods of AFM is to enhance the imaging capabilities of AFM or recognize patterns on a surface, one could consider the use of statistical methods \cite{0957-4484-20-8-085714,0957-4484-20-40-405708}  as a means of revealing dependencies in the data set of intermodulation amplitude and phases measured at each image pixel.   One can also simply plot surface maps of the intermodulation amplitudes and phases \cite{platz_ultramicroscopy} to discover if new features arise in the image which were not visible in standard dynamic AFM.  Nevertheless, a physical understanding of the origin of the newly observed features is desirable.   To this end one can resort to direct numerical simulation of the dynamical system with an appropriate model for the tip-surface force, as a means of understanding how parameters in a force model effect the intermodulation response.

Here we report on the results of numerical simulations using a single eigenmode cantilever and conservative tip-surface force model, where the Young's modulus of the surface is varied over six orders of magnitude.  We simulate the response as the oscillating cantilever approaches a surface, and compare the results with experimental approach curves taken on three different materials spanning this range of Young's moduli.  In both simulation and experiment we calculate and measure respectively, both the amplitude and phase of the response at intermodulation frequencies while approaching the surface.  Comparison between the simulation and experiments allow us to draw some qualitative conclusions about the origin of characteristic features in the approach curves.  In particular, we find that the higher order intermodulation response results from stiffer surfaces, and that the phase of higher order intermodulation products is quite responsive to small changes in the approach distance, making this signal very suitable for feedback control of the probe height.  

\section{Numerical simulations}

We model the cantilever dynamics with a single eigenmode, for example the fundamental bending mode of the cantilever \cite{Raman200820}.  This approximation will be valid as long as the eigenmodes have a sharp resonances, and the drive and response have significant frequency components only close to one eigenmode. The single eigenmode approximation allows us to treat the cantilever as a simple harmonic oscillator with a linear restoring force in $z-z_0$ , where $z$ is the the tip-surface separation, and $z_0$ is the tip location when the cantilever is in it's equilibrium position.  

\begin{equation}
\ddot z +\frac{1}{Q} \dot z + (z-z_0) = \frac{1}{k} (F_{\mathrm{drive}}+F_{\mathrm{ts}})
\label{eq_sho}
\end{equation}
Here $\dot z $ means differentiation with respect to dimensionless time $\tau=2 \pi f_0 t$.  The simulation used values of the quality factor $Q = 510$, resonance $f_0 =277$~kHz, and spring constant $k=28$~N/ms, which are typical for the experiments described in the next section. 

The nonlinear tip-surface force is modeled in a piece-wise fashion using the van der Waals - DMT model \cite{Garcia2002197}.  In this model the tip is approximated by a sphere of radius $R$ which is attracted toward a planar surface of uniform composition by the van der Waals force.  The attractive force is cut off at $z=a_0$, where the model switches to a repulsive contact force due to the mutual elastic deformation of a spherical tip and planar surface,  

\begin{equation}
F_{ts}(z)=\left\{ 
\begin{array}{ll}
-\frac{HR}{6z^{2}} & \mathrm{for~} z > a_0\\
-\frac{HR}{6a_{0}^{2}}+\frac{4}{3}E^{*}\sqrt{R}\left(a_{0}-z\right)^{3/2}
& \mathrm{for~} z \leq a_0
\end{array}\right.
\end{equation}
The Hamaker constant $H$, and the cut-off distance $a_0$ are the parameters controlling the attractive van-der Walls force, and the effective modulus $E^*$ characterizes the repulsive contact force. 
\begin{equation}
\frac{1}{E^*}=
\frac{1-\nu_{\mathrm{tip}}^2}{E_\mathrm{tip}}+
\frac{1-\nu_{\mathrm{surface}}^2}{E_\mathrm{surface}}
\end{equation}

\begin{figure}[ht]
\begin{center}
\includegraphics[width=8cm]{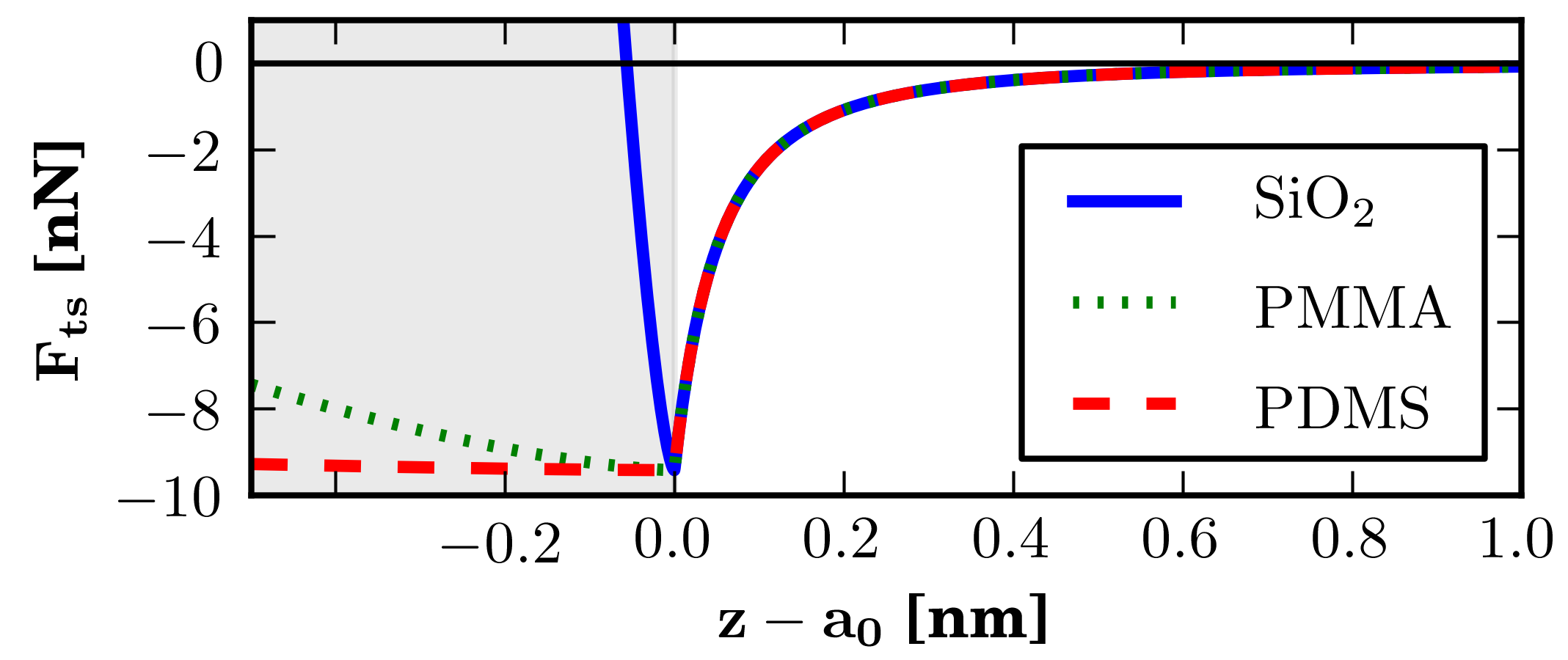}
\end{center}
\caption{THE MODEL USED FOR THE TIP-SURFACE FORCE IN THE NUMERICAL SIMULATIONS, SHOWN FOR THREE DIFFERENT VALUES OF THE YOUNG'S MODULUS OF THE SURFACE.}
\label{fig_force} 
\end{figure}

For the sake of limiting parameter space in the simulation, we keep the attractive force constant by fixing  $H=6.0$ x $10^{-20}$~J, $a_0=0.103$~nm and $R=10$~nm, and we neglect differences in the Poisson ratios by fixing $\nu_{\mathrm{tip}}=\nu_{\mathrm{surface}}=0.53$. The stiffness of the surface is then varied by changing the Young's modulus of the surface over six orders of magnitude.   Representative tip-surface force curves are plotted in fig. \ref{fig_force} for a Si tip $E_{\mathrm{Si}}=120$~GPa, and three different values of $E_{\mathrm{surface}}$ corresponding to the materials studied in the experimental section  $E_{\mathrm{SiO2}}=70$~GPa, $E_{\mathrm{PMMA}}=1.2$~GPa and $E_{\mathrm{PDMS}}=50$~kPa.   

\begin{figure*}[ht]
\begin{center}
\includegraphics[width=17cm]{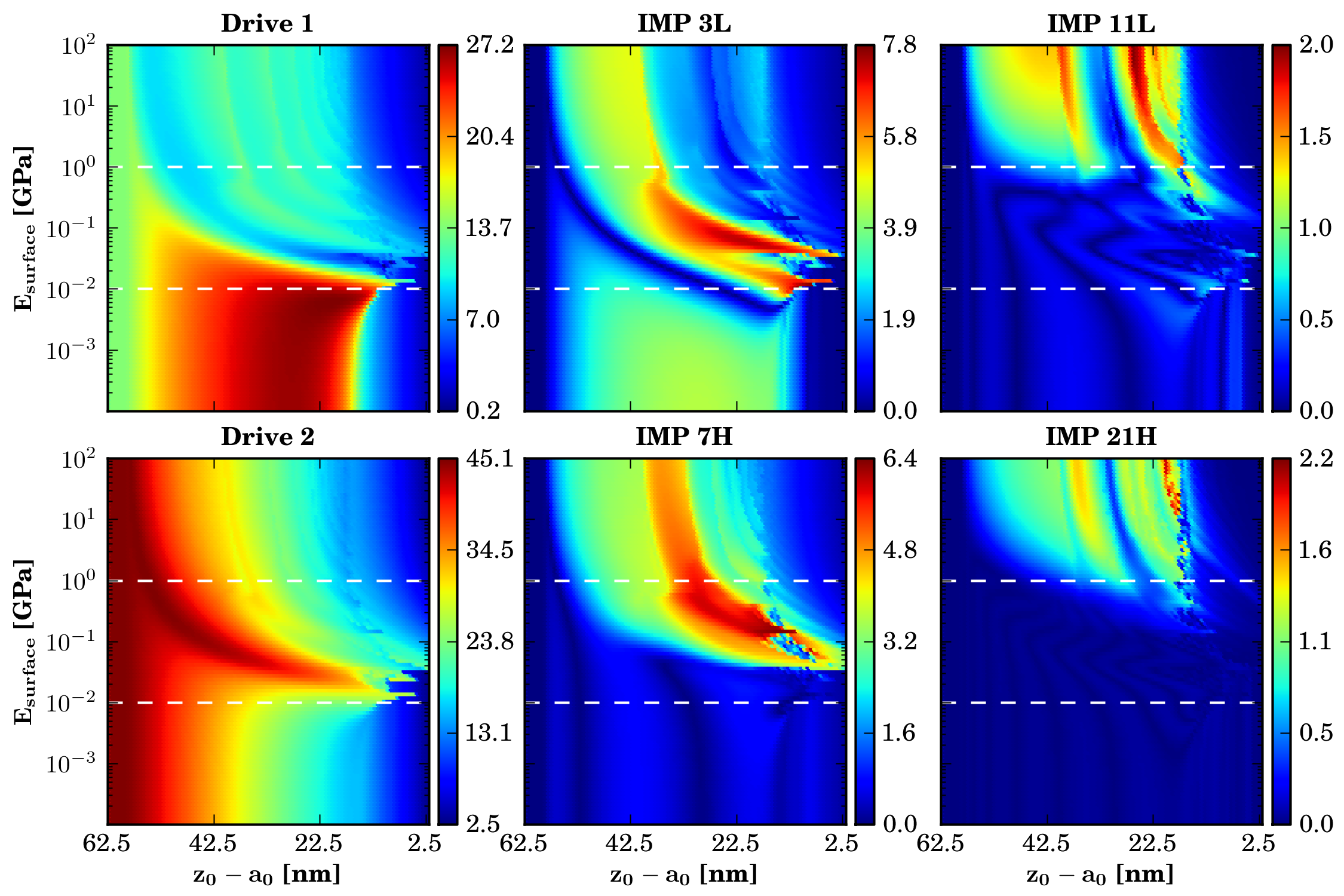}
\end{center}
\caption{COLOR MAPS SHOWING THE SIMULATED RESPONSE AMPLITUDE AT THE DRIVE FREQUENCIES AND FOUR OF THE MANY INTERMODULATION PRODUCTS, IN THE PLANE OF THE YOUNG'S MODULUS OF THE SURFACE VS. THE APPROACH DISTANCE.}
\label{fig_map_amp} 
\end{figure*}

The drive force consists of two pure harmonic tones,
\begin{equation}
F_{\mathrm{drive}}=A_1 \cos(\Omega_1 \tau) + A_2 \cos(\Omega_2 \tau)
\end{equation}
at the drive frequencies $\Omega_{1,2}=f_{1,2}/f_0$ which have a greatest common divisor $\Delta \Omega$.  All intermodulation products appear at frequencies which are integer multiples of $\Delta \Omega$, so the response can be expressed as a Fourier series in $\Delta \Omega$ \cite{Hutter_PRL}.  In the simulations one drive was placed slightly below resonance, $f_1=276.5$~kHz and the other drive on resonance $f_2=277$~kHz, so that $\Delta \Omega / 2 \pi =f_2 - f_1$ = 500 Hz. We note however, that there are several possible drive configurations which produce many intermodulation products near resonance.  

The integration of equation (\ref{eq_sho}) was done numerically using the solver CVODE contained in the SUNDIALS suite of nonlinear solvers \cite{1089020}.  CVODE is a variable-order, variable-step integrator with a built in root finding, or discrete event detection routine, used here to ensure that the solver generates a discrete output at $z=a_0$.  We experimented with the two families of multi-step methods provided in CVODE, Adams Moulton Formulas and Backward Differential Formulas, which are recommended for non-stiff and stiff problems respectively.  In both cases, the functional iteration method was used.  We found no discernible difference between these various methods for the study reported here.  The output of the integrator is sampled in time and Fast Fourier Transformed to get the response spectrum.  Care is taken to chose a sampling frequency which is an integer multiple of $\Delta \Omega$, so that points in the discrete Fourier transform land exactly at intermodulation frequencies.

In order to explore the effect of surface stiffness on the response, we fix the cantilever spring constant to $k = 28\ \mathrm{N/m}$ and the values of the $H$, $a_0$, and $R$ as given for fig. \ref{fig_force} , and simulate the response as we step $z_0$ toward the surface.  This approach simulation is repeated in a loop stepping the effective modulus $E^*$, logarithmically over six orders of magnitude.  We generate color density plots of the response amplitude in the parameter plane of $E^*$ vs. $z_0$.  Figure \ref{fig_map_amp} shows six such plots, at each of the two drive frequencies $\Omega_1$ and $\Omega_2$, and four of the intermodulation frequencies: the third order intermodulation product 3L at frequency $\Omega_1 - \Delta \Omega$, the 11$^{\mathrm{th}}$ order intermodulation product 11L at $\Omega_1 - 5 \Delta \Omega$, the 7$^{\mathrm{th}}$ order intermodulation product 7H at frequency $\Omega_2 + 3 \Delta \Omega$, and the 21$^{\mathrm{st}}$ order intermodulation product 21H at $\Omega_2 + 10 \Delta \Omega$.  

These color maps show that the intermodulation response is rich and varied over the parameter space explored.  Nevertheless, we can observe some general trends which are best described by comparing three regions of surface stiffness, as indicated by the horizontal dashed lines in fig. \ref{fig_map_amp}.  

At low surface stiffness, below $10^{-2}$ GPa, we see that the response at the drive frequencies changes very little with stiffness.  In the simulation drive 1 and drive 2 have equal strength, but the response amplitude of the free cantilever (left edge of the panels of fig. \ref{fig_map_amp}) is lower at drive 1 than drive 2.  This is because drive 2 is on resonance, and drive 1 is off resonance by 500~Hz, below drive 2.  When engaging the surface, we see that the response at drive 1 increases in amplitude during the initial approach, whereas the response at drive 2 decreases.  We can understand the relative change of the drive amplitudes as resulting from a parameter change of a linear system: the attractive tip-surface force effectively weakens the linear cantilever restoring force, causing a shift of resonance toward lower frequency, away from drive 2, toward drive 1.  However, this naive approach neglects the redistribution of power between frequencies, forbidden in linear systems, but characteristic of nonlinear systems.  Upon approaching the surface, we also see the appearance of intermodulation products of the two drives.  Response at these frequencies can only be understood by considering the nonlinear response of the system.  For the nonlinear system, power can be taken from one drive and transfered to the other drive (amplification), and to the intermodulation products of the two drives.   

In this low stiffness region we also see that intermodulation products of high order have a low response amplitude, which shows an oscillating behavior as the surface is approached.  The order of the intermodulation product roughly corresponds to the order of the term in a power series approximation to the tip-surface force \cite{Hutter_PRL}.  For low stiffness, coefficients of high powers of $z$ in a polynomial approximation of $F_{\mathrm{ts}}(z)$, will be small in comparison to those for high stiffness (see fig. \ref{fig_force}).  Thus, a rule of thumb is: the stiffer the surface, the larger the response of high-order intermodulation products.

At intermediate stiffness, between $10^{-2}$ and $10^{0}$ GPa, we see a rapid drop in the response amplitude at drive 2, and a corresponding increase in the amplitude at drive 1.  We also observe a peaking of the intermodulation response amplitudes 3H, and at somewhat higher stiffnesses, 7H. In all of the aforementioned, we observe sharp tongues of high amplitude extending toward smaller approach distance, where they become unstable, switching from low to high amplitude. At such a close approach and large amplitude oscillation, the nonlinearity is too strong and a bifurcation of the dynamical system will occur \cite{1461509,Garcia2002197,hashemi:093512} resulting in bi-stable or multi-stable oscillation states.  Bifurcations cause unstable imaging conditions because unavoidable noise causes jumps between the resulting meta-stable oscillations states.  It is therefore reassuring to see that at moderate approach distances of 22~nm and above, we find stable behavior over a wide range of surface stiffness.  Indeed, experiments on multi-frequency AFM report more stable behavior than standard single-drive dynamic AFM for comparable operation conditions \cite{Platz2008,thota:093108}.

Finally, at high stiffness, above $10^{0}$ GPa, we find a sharp reduction in the response amplitude at drive 1 coinciding with a sharp increase in the amplitude of the 3$^{rd}$ order intermodulation product, as drive power is redistributed.  Proceeding to higher stiffness, we see that the drives and lower order intermodulation response show little change, whereas higher order intermodulation response show greater variation with stiffness.  Again, this is to be expected because for very stiff surfaces, the polynomial approximation of $F_{ts}(z)$ will have significant contributions from high powers of $z$.  These parameter maps, together with those for other intermodulation products, show that intermodulation is capable of producing a varied response over a very wide range of surface stiffness, for one value of the cantilever spring constant.  This complex behavior of the various intermodulation products which is observed upon approaching the surface, demonstrates that spectrum of intermodulation response at one approach distance provides excellent fingerprint of the material and mechanical surface properties at the nanometer scale.

\section{Experiments}

\begin{figure*}[ht]
\begin{center}
\includegraphics[width=17cm]{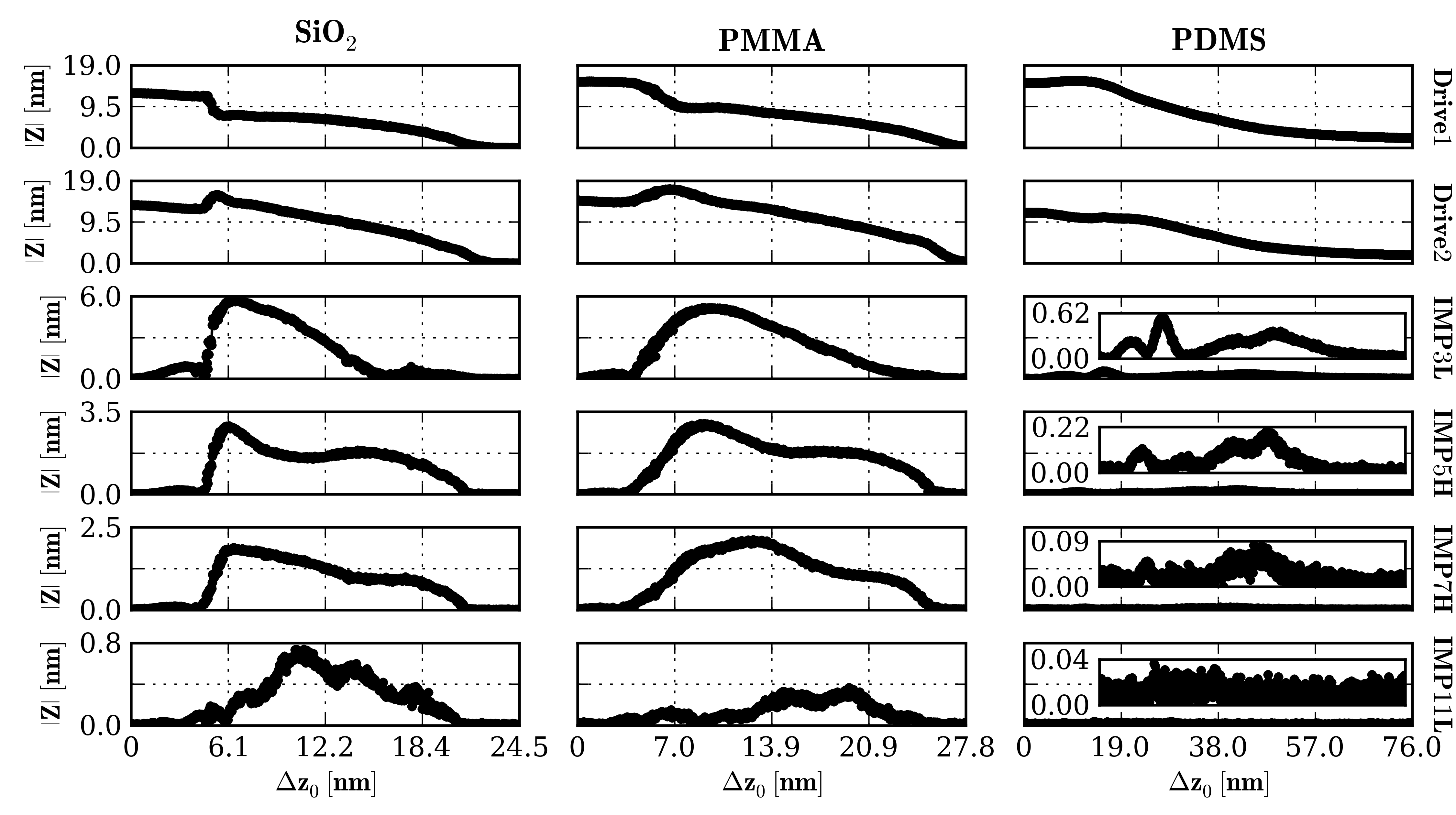}
\end{center}
\caption{THE AMPLITUDE OF RESPONSE $Z$ AT THE TWO DRIVE FREQUENCIES AND FOUR INTERMODULATION PRODUCTS AS A FUNCTION OF CHANGE IN APPROACH DISTANCE FOR THREE MATERIALS OF WIDELY VARYING YOUNG'S MODULUS.  THE VERTICAL AXES IN EACH ROW ARE IDENTICAL AND THE INSETS SHOW A VERTICAL ZOOM WHERE THE RESPONSE WAS WEAK.}
\label{fig_experiments} 
\end{figure*}

\begin{figure*}[ht]
\begin{center}
\includegraphics[width=17cm]{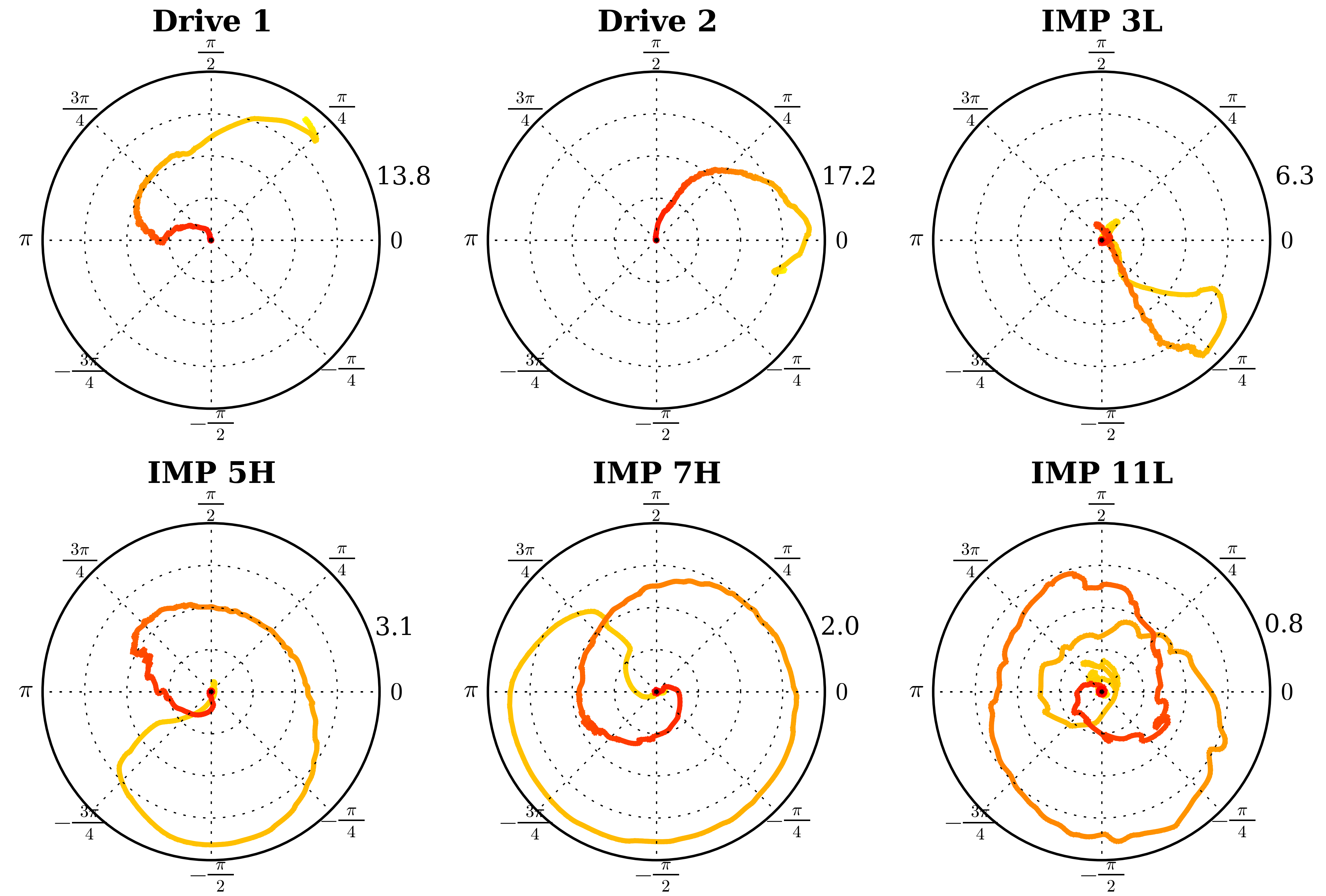}
\end{center}
\caption{PHASOR PLOTS SHOWING HOW THE AMPLITUDE AND PHASE OF THE RESPONSE EVOLVES AS THE PROBE APPROACHES AN SiO$_2$ SURFACE. THE NUMBERS GIVEN JUST OUTSIDE THE FIRST OCTANT REFER TO THE AMPLITUDE AT FULL SCALE (nm). THE COLOR GRADIENT CODES APPROACH DISTANCE, WHERE RED CORRESPONDS TO CLOSEST APPROACH.}
\label{fig_polar} 
\end{figure*} 

Experiments were performed in air on a Veeco Multimode~2 AFM, with additional electronics for synthesizing the drive signal and a separate computer for sampling and analyzing the cantilever response \cite{platz_ultramicroscopy}.  Cantilevers of the type MP-11100-10 from Veeco were used in the experiment.  The resonant frequency and quality factor were determined by measuring the thermal equilibrium fluctuation force due to the damping medium (air), which can be observed near resonance where sensitivity is enhanced.  Typical values for these cantilevers were: $Q=575$ and $f_0=290$~kHz.  By calculating the hydrodynamic damping from the cantilever dimensions and density of air \cite{sader:3967,Higgins2006}, we can determine both the spring constant of the cantilever and the optical lever responsivity of the AFM, without touching the surface. Typical values for these experiments were $k=28$~N/m and $\alpha = 55$~nm/V respectively.  

When acquiring the approach curves, the Veeco system is set to make a ramp toward the surface with ramp frequency 0.1~Hz, while a sampling card streams data to storage at a sampling frequency of 2~MHz.  The stored data is parsed and Fourier transformed to capture the intermodulation spectrum at different approach distances.  In the experiment we can only control the change in approach distance $\Delta z_0=z_{\mathrm{start}}-z_0$, where the origin of the approach $z_{\mathrm{start}}$ is chosen to be at the onset of intermodulation response. Measurements were made approaching the surface, and immediately retracting, taking care not to go too far beyond the point where all response amplitudes extinguish. 
By overlaying the approach and retract data, we found that the data fell on the same curve, with no visible sign of hysteresis.  This indicates that the oscillating cantilever could not be trapped in differing, meta-stable oscillation states.  Taking care to avoid tip and surface damage by not approaching too close to the surface, we were able to get quantitatively consistent results for fixed drive parameters, when comparing several consecutive measurements at one point on a surface.  When comparing the response at several different points on the same surface, we also found quantitatively self-consistent results for each surface studied.  Measurements were preformed on each surface with two different cantilevers, where comparison showed a qualitative self-consistency.  The small differences observed between different cantilevers may be explained by variation in the cantilever parameters and the placement of the drive frequencies with respect to resonance.  

Representative curves showing the amplitude of the response at the two drive frequencies and four of the intermodulation products are show in fig. \ref{fig_experiments}, for approach toward three different surfaces:  The SiO$_2$ surface was a piece of a Si wafer with 2~$\mu$m thermal oxide on the surface.  The PMMA (molecular weight 950 kDalton) was spin-coated on a piece of Si wafer to a thickness of about 1.3~$\mu$m.  The PDMS surface was cast on to a glass surface and vacuum cured to form a sheet 1~mm thick.  This sheet was then piled off and placed on an AFM chuck so that the surface which cured against the glass could be probed.

The experimental curves of fig. \ref{fig_experiments} reveal some interesting similarities and differences from the simulated response.  Similar to what was expected from the simulations, we find that the amplitude of higher order intermodulation products is much weaker for the softest material PDMS, in comparison with either PMMA or SiO$_2$.  In contrast to what we expected from the simulation, we find that PMMA and SiO$_2$ have qualitatively similar approach curves.  Upon first contact with the surface, each material shows a small initial maximum in the response amplitude of the low-order intermodulation products.  This initial maximum is also seen in the simulations at intermediate stiffnesses.  However, judging from the relative size of this feature in the two materials, comparison with simulation (see fig. \ref{fig_map_amp} IMP 3L) would indicate that SiO$_2$ is stiffer than PMMA, which is clearly not the case.  

From our experience with other simulations, we find that this initial maximum is sensitive to both attractive and dissipative forces.  The former was kept constant and the latter was absent in the simulations.  We modeled the tip-surface force with the van der Waals - DMT model, changing only the surface Young's Modulus.  This conservative force model does not account for dissipative effects which may be caused by adsorbed water molecules on the surface.  The fact that response from SiO$_2$ appears softer than expected from our simulations could be explained by the presence of a surface adsorbate on the SiO$_2$.  

Comparing the higher order intermodulation amplitudes of  SiO$_2$ and PMMA in fig. \ref{fig_experiments} we again find qualitative consistency with the simulations.  The stiffer SiO$_2$ surface generates a larger amplitude of response at the 11$^{th}$ order intermodulation product IMP 11L, than does the softer PMMA surface.  Overall we see a striking qualitative similarity in features of the experimental amplitude vs. $\Delta z_0$ curves of SiO$_2$ and PMMA at each frequency, where features in the PMMA curves appear smother than the corresponding features in the SiO$_2$ curves.

\section{Intermodulation phase}

The previous sections discussed only the amplitude of the response at the drive frequencies and a few of the many intermodulation products.  It is also possible to determine the phase of the response at all of these frequencies because they are integer multiples of a fundamental frequency in the problem, $\Delta \Omega$, which is the greatest common divisor of the two drive frequencies.  That all response can be accounted for by a Fourier series in $\Delta \Omega$ is a valid assumption as long as the nonlinearity is not too strong, so that period doubling or other such precursors to chaos do not appear in the response.  In the experiment, the phase can be determined by using the two drives signals to build a reference signal with frequency $\Delta \Omega$ \cite{platz_ultramicroscopy}.  This measurement constituents a generalization to nonlinear systems, of a common instrument used in linear systems analysis, known as the lock-in amplifier, or network analyzer.

To represent both the amplitude and phase of the response, it is convenient to plot the approach curves in a polar coordinate system  as seen in fig. \ref{fig_polar}.  Each point in the plot corresponds to the response at a particular value of $\Delta z_0$.  A vector stemming from the origin to this point (a phasor) has length which is the amplitude, and polar angle which is the phase.   The response at the two drive frequencies starts at the high amplitude of the freely oscillating cantilever and evolves toward zero along a contorted path as the cantilever approaches the surface.  Response at the intermodulation frequencies starts at the origin of the polar plots because the freely oscillating cantilever is a linear system with no intermodulation response.  When the surface is engaged the intermodulation response appears and we observe much greater variation of the phase, where the path loops several times around zero for higher order intermodulation products.

\begin{figure}[ht]
\begin{center}
\includegraphics[width=8cm]{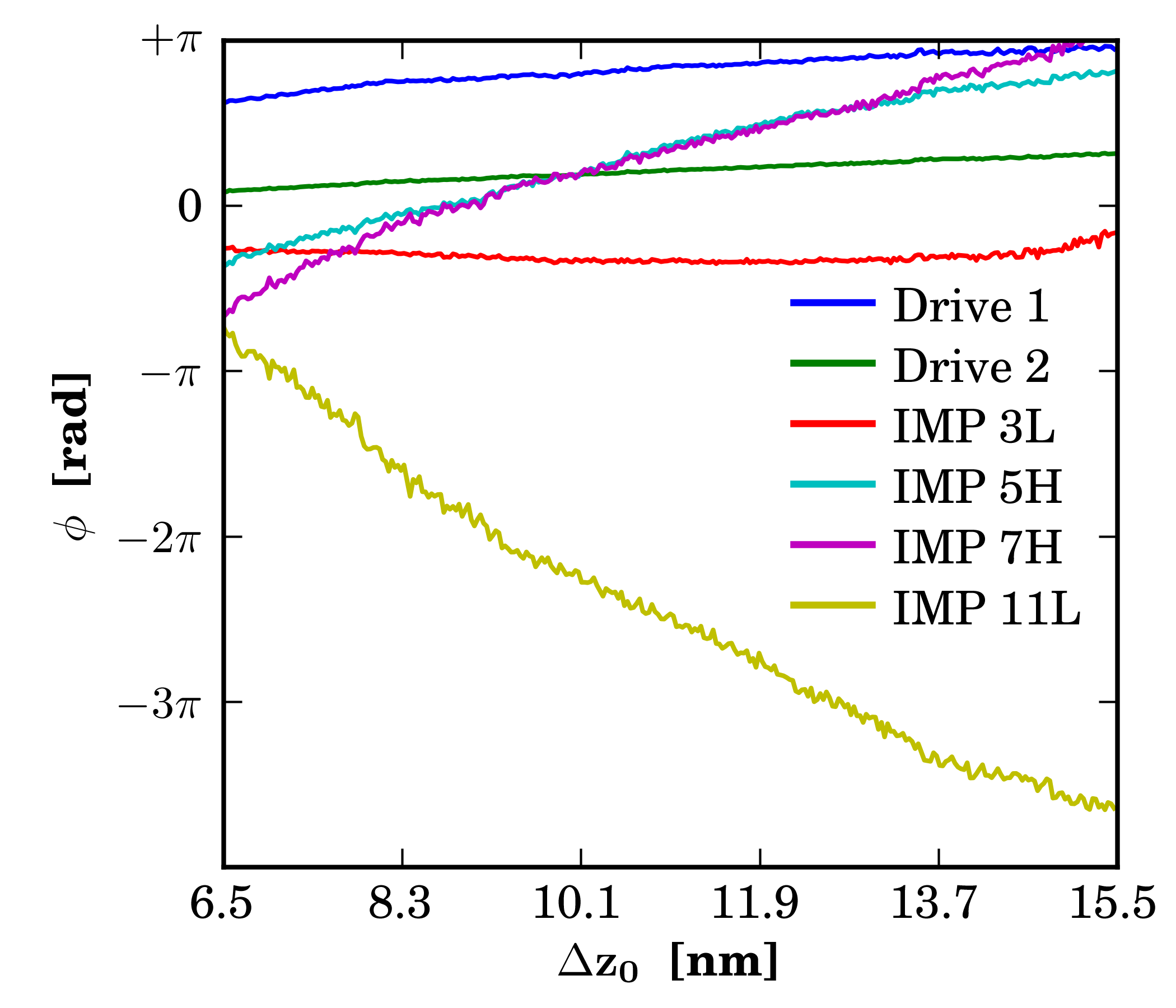}
\end{center}
\caption{THE PHASE OF THE RESPONSE $\phi$, PLOTTED AS AN EXTENDED VARIABLE, VS. THE APPROACH DISTANCE.}
\label{fig_phase} 
\end{figure}

This multiple looping of the higher order intermodulation products means that the phase is winding more rapidly for higher order.  Indeed, if we unwind the phase during the approach, we can plot the phase as an extended variable over an interval of several times $2 \pi$ as the surface is approached.  Figure \ref{fig_phase} shows such a plot for the same curves as those given in fig. \ref{fig_polar}.  Here we clearly show how intermodulation products at frequencies higher than the drive advance in phase, whereas those lower than the drive retard in phase, upon approaching the surface (winding in opposite sense in fig. \ref{fig_polar}).  We also see that the phase for higher order intermodulation products is much more responsive to changes in $\Delta z_0$ than the phase at either drive frequency, indicating that the higher order phase would make a good feedback signal for controlling the AFM probe height.  

\section{CONCLUSION}

We have described the amplitude and phase response due to the intermodulation of two pure harmonic drive tones in dynamic AFM, showing how these quantities change upon approach toward a surface.  Simulations of the approach process with a single eigenmode model using the van der Waals - DMT, conservative, nonlinear tip-surface force, show that the intermodulation response is rich and varied over a wide range of surface stiffnesses.  Experiments on three surfaces spanning this range of Young's modulus of the surface also show rich and varied response.  However, it is difficult to correlate the experimental and simulated results in any quantitative way.  These difficulties stem from the use of an idealized tip-surface model in the simulation.  The model is not expected to be accurate for softer surfaces, and it neglects the dissipative processes due to an absorbed water layer, which can arise in AFM performed in ambient air at standard temperature and pressure.  Furthermore, the model assumes an ideal geometry of a round tip and a flat, semi-infinite surface, both being homogeneous in composition.  While the experiments here probed flat, homogeneous surfaces, samples of interest will often have considerable variation in topography and composition of the surface at the nanometer scale.  Because intermodulation AFM is very sensitive to small changes in the nonlinear tip-surface force, we expect from theory that the observed intermodulation response will also be strongly effected by the  topography and local composition of the surface.  Thus, for real samples, the idealized force model used here is not particularly useful.  Indeed, bulk elastic moduli or Hamaker constants for a specified geometry, are not the proper quantities for characterization of nanostrucutred surfaces.  Exactly which quantities best characterize the surface properties probed by dynamic AFM, is an open question for AFM research.  It is the belief of the authors that intermodulation AFM can play an important role finding the answer.

\section{Acknowledgments}
We gratefully acknowledge funding from the Swedish Research Council VR, Vinnova and G\"oran Gustafsson Foundation.

\end{document}